\input harvmac

\def\np#1#2#3{{\it Nucl. Phys.} {\bf B{#1}} ({#2}) {#3}}
\def\pl#1#2#3{{\it Phys. Lett.} {\bf B{#1}} ({#2}) {#3}}

\def\hep#1{{\bf hep-th/{#1}}}

\def\alg#1{{\bf alg-geom/{#1}}}

\Title{hep-th/9611007, HUTP-96/A049, IP/BBSR/96-48}
{\vbox{\centerline{U-Manifolds}}}
\centerline{Alok Kumar}
 \medskip\centerline{\it Institute of Physics}
 \centerline{\it Bhubaneswar 751 005, INDIA}
\vskip 0.15in
\centerline{and}
\centerline{Cumrun Vafa }
\medskip\centerline{\it Lyman Laboratory of Physics}
\centerline{\it Harvard University}
\centerline{\it Cambridge, MA 02138, USA}
\vskip 0.15in
\vskip .3in
\noindent
We use non-perturbative U-duality symmetries of type II strings
to construct new vacuum solutions.  In some ways
this generalizes the F-theory vacuum
constructions.  We find the possibilities
of new vacuum constructions are very limited.
Among them we construct new theories with $N=2$ 
supersymmetry in 3-dimensions and $(1, 1)$ 
supersymmetry in 2-dimensions.
\Date{1/97}
\newsec{Introduction}
It has now been appreciated that in many cases string theories enjoy
more symmetries when quantum corrections are taken into account.
A particular case of this is the enlargment of T-duality symmetries
of type II strings to U-duality symmetries
\ref\ht{{C. Hull and P. Townsend, ``Unity of Superstring Dualities'',}
{\np{438}{1995}{109}}.}\ref\wid{{E. Witten,
``String Theory Dynamics in Various
Dimensions",} {\np{443}{1995}{85}}.}.
A special case of this involves type IIB in 10 dimensions which
is conjectured to have $SL(2,{\bf Z})$ U-duality symmetry, or type IIA
compactified on a circle which again is
conjectured to have the $SL(2,{\bf Z})$ U-duality
symmetry.  Symmetries often have a geometric origin.  It is
natural to look for a way to `geometrize' U-duality symmetries.
For type IIA compactification
to 9 dimensions this is done \ref\sch{{J. Schwarz, ``An SL(2, Z)
Multiplet of Type IIB Superstrings'',}
{\pl{360}{1995}{13}}.}\ref\asp{{P. Aspinwall,
``Some Relationships between Dualities
in String Theory'',} {\hep{9508154}}.}\
by viewing the theory as coming from the 11 dimensional M-theory
compactified on $T^2$ to 9 dimensions,
where the $SL(2,{\bf Z})$ duality gets
identified with the duality symmetries of the internal $T^2$.
A similar motivation in geometrizing the U-duality of type IIB
leads to the notion of a 12 dimensional F-theory
\ref\vf{{C. Vafa, ``Evidence for F-theory'',} {\np{469}{1996}{403} }.}\
whose compactification on $T^2$ by definition gives
rise to type IIB in 10 dimensions.

Quite aside from fundamental issues as to the physical meaning
of such geometrizations,
the power of geometrizing U-dualities lies in the fact that
new vacuum solutions can be effectively
 constructed in this setup.
If we allow
the scalar fields (the `U-fields') to vary over space and allow
them to jump, consistent with U-dualities, this data
will translate to a manifold whose base is the `visible' space
and the fiber being the geometrization of the U-duality.
Questions of smoothness of compactifications
are much easier to address in this geometric framework.
In particular a criterion
for having a `good' vacuum is that the total
space consisting of the base and the auxilliary space which
is the fiber be smooth.

More can be said about usefulness of geometrization
of U-dualities when the supercharges transform according
to the spinor of both the uncompactified space as well
as the internal fiber of the U-space.  In particular
the question of the number of supersymmetries
one preserves gets mapped to the number of covariantly constant
spinors on the {\it total space}; this is a subject which
is of course well studied.

The vacuum constructions in F-theory can be viewed as utitilizing
the $SL(2)$ symmetry of type IIB in 10 dimensions.  Similarly,
compactifications of M-theory on elliptic manifolds can be viewed
as geometrizing the U-duality of type II strings (again $SL(2)$)
in 9-dimensions.
Given the power of M- and F-theory in constructing new vacua,
it is natural to ask if one can use the U-dualities more generally
to construct additional new vacua.  The aim of 
this paper is to take a first
step in this direction.  Even though we will be able
to construct some new vacua, in some sense the possibilities appear
 to be far more limited than in the case of M- or F-theory.

\newsec{General setup}
The moduli space of vacua for a string theory with
enough supersymmetry and in particular for toroidal
compactification of type II strings typically
involves a coset space
$$ {\hat G}/G(Z)\times H$$
where $H$ is the maximal compact subgroup of ${\hat G}$, and $G(Z)$
is the  U-duality group which one identifies
as discrete gauge symmetry of the theory. The charges
form representations of $\hat G$.  This is also true for supercharges,
but in that case by an appropriate redefinition of the fields
we can take them to transform only under the compact subgroup
$H\in \hat G$.
  In the following
we will abbreviate $G(Z)$ by writing it simply as $G$. Once
we fix a point $p$ in the moduli space of the theory by setting
the scalars to some expectation value, there will in general
be some subgroup  $G_p\subset G$ which preserves that point.
We identify $G_p$ as part of the symmetry of that particular background.

We can contemplate a number of
ways to use this symmetry to construct new vacua:

1) Modding out by a subgroup $K\subset G_p$.  This is the generalization
of the orbifold idea and includes the standard orbifold
compactifications as a special case.
In particular in the case of compactifications
of type II string on a $d$-dimensional torus, we have the perturbative
symmetry $G_p\subset SO(d,d;Z)$.  Constructions of orbifolds, symmetric as
well as asymmetric ones, involve a choice of a subgroup
of $G_p$.  We can accompany the group action with a phase action
depending on the charged states in the theory.  
In particular say the charges
form a $D$-dimensional representation of $\hat G$.  Let $a$ be
a vector in $R^D$.  We can enlarge the group we mod out
by including some action on the charged states according to
$$(\theta , a)|Q\rangle 
\rightarrow {\rm exp}(i a\cdot Q)|\theta Q \rangle$$
where $\theta $ denotes
the action of the orbifold group element on the charge lattice.  This
generalization
is also familiar
from orbifold constructions where one introduces Wilson line.

2)  We can further compactify on another space,
say a d-dimensional torus and mod out by  a symmetry
which partly acts on the torus and partly on the internal U-duality
symmetry $H$.  This can of couse be reduced to the case 1)
by considering the U-duality as to include
that one obtains upon compactification on $T^d$,
but will be useful to keep it as a separate case for reasons
we shall explain later. Some examples of such orbifolds, 
corresponding to the $SL(2, Z)$ S-duality symmetry of the 
type IIB theories, have been presented in 
\ref\md{{K. Dasgupta and S. Mukhi, ``F-theory at Constant 
Coupling'',} {\hep{9606044}}.}.

3) We can consider the case where the fiber fields
are varying smoothly over the space (base) except for some loci
of singularities around which the fiber undergoes monodromies
belonging to $H$.  This case can be viewed as generalizing case
2) above.

The analogy with orbifold constructions is very helpful and
points to some subtleties that have to be overcome.  In discussing
these theories we will use the terminology of orbifold constructions
and in particular the notion of twisted sectors.

In cases of constructions 1) and 2) it is straightforward
to deduce the massless modes which survive the projection in the
untwisted sector.  The difficulty lies in finding the spectrum and
interactions involving
the twisted states.   In case 3)  in general there is
no division of states to twisted and untwisted sectors
and we will have to find another
way to find the low energy lagrangian.

The basic strategy in determining the spectrum we have
is as follows:  Consider further compactification on $T^d$
and assume that for some $d$, upon
conjugation by an element of the $U$-duality group,
$G$ can be viewed as a subgroup of the T-duality group in the
compactified theory.
In this way, assuming U-duality conjugation commutes
with orbifolding, we can deduce the states of the twisted
sectors using the standard orbifold techniques for cases 1) and 2)
and using Kaluza-Klein
compactification techniques to deduce the spectrum in case 3).
 Of course we may lose
some information about the massless modes 
in the higher dimensional theory
(such as chirality)
with this procedure, but very often some knowledge of supersymmetry
together with absence of anomalies can be used 
to reconstruct the massless
modes of the original theory.

This trick of deducing massless modes upon dimensional reduction
is not reliable for case 1) because in such cases there is evidence
that U-duality and orbifolding does not commute; see for example
\ref\vw{{C. Vafa and E. Witten, ``Dual String Pairs with
N=1 and N=2 Supersymmetry in Four Dimensions,''} {\hep{9507050}}.}.
However in cases 2) and 3)
it appears to be reliable, as has been checked in many
cases in
\ref\sen{{A. Sen, ``Duality and Orbifolds'',} {\hep{9604070}}.}.
There is a good reason why
in cases 2) and 3) it should have worked: In such case
the twisted sectors (or the loci where
the fiber becomes singular)  can be viewed as p-branes
of the higher dimensional theory and in this sense the
consistency of conjugating with U-duality amounts to checking
the transformation property of the twisted 
sector p-branes under U-duality.
This can thus be viewed as a check of the U-duality in the higher
dimensional theory.
Note that the fact that we cannot determine the spectrum
in case 1) reliably does not imply that orbifolding 
does not make sense.
It is simply a reflection of lack of a technique to 
deduce its properties.
For this reason we will mainly concentrate on cases 2) and 3) 
in this paper.

In order to effectively use the geometrization
of the U-duality symmetries we need the compact
part of the U-duality be a subgroup of some holonomoy
group and the fiber needed to construct the U-manifold
is motivated by how the compact part of the U-duality group
is to act on the fiber (recall that the supercharges
transform according to a representation of $H$).
  For simplicity in this paper we will deal
with the case where the compact part of the U-duality group $H$ is
 $\otimes_i SO(n_i)$ for some $n_i$ and that the
supersymmetry charges transform in some spinor representation
of $\otimes_i SO(n_i)$ and in particular  we will 
restrict our attention in
this paper to the U-dualities
which arise upon toroidal compactification of type II strings
to $d\geq 6$ spacetime dimensions.  The basic idea of this paper
clearly generalizes to the other cases as well.  In fact an
interesting case to consider may be the case with affine U-duality
symmetry when we compactify down to
$d=2$ \ref\jul{{B. Julia, ``Group Disintegrations'',}
{in {\it
Superspace and Supergravity}, Eds. S. Hawking and M. Rocek,
(Cambridge University Press , 1981) 331-349}.}.

  The auxilliary space we will use for the fiber
of our construction will involve $\otimes_i T^{n_i}$.
In this way when the base manifold
uses some holonomies in the $\otimes SO(n_i)$ we can determine
how much supersymmetry it preserves.

\newsec{U-dualities}
Let us start with U-dualities in higher dimensions 
and go down in dimension.
We will limit ourselves in this paper to the U-dualities for type II
compactifications to 6,7,8,9 and the uncompactified 
10 dimensional case.

\subsec{Dimension 10 and type IIB}
In this case $H=SO(2)$.  We can attach an auxilliary $T^2$.
The supercharges transform in the spinor of $SO(2)$ of a given
chirality:
$$Q=({\bf S}_{10}^+\otimes {\bf 1}_s^+)
\oplus ({\bf S}_{10}^+\otimes {\bf
1}_s^-)$$
where ${\bf S}_{10}^+$ denotes the spinor of positive chirality
of $SO(9,1)$ and ${\bf 1}_s^{\pm}$ denote spinors of $SO(2)$
of $\pm$ chirality.
If we go down on another $T^2$, this $SO(2)$ is contained, by
$U$-duality conjugation, in the T-duality of string theory $SO(2,2)$.
In particular it is conjugate to complex structure symmetry
of type IIA on $T^2$. In fact, in this case we do not have
to go all the way to string theory to find the structure
of this theory:  upon compactification on $S^1$, this $SO(2)$
can be identified with the symmetry of the $M$-theory compactification
on $T^2$, and so we get a connection with M-theory which at
least for smooth compactifications of type 3) can be used to
give the massless states.  This chain of dualities implies that
if we have a manifold $K$ with $T^2$ fibers (with a section) we
can construct new vacua.  This was the original motivation
for the introduction of F-theory \vf .  Moreover the embedding
of the U-dualities mentioned above implies that considering
F-theory on $K\times S^1$ is the same as considering M-theory
on $K$.  Moreover considering F-theory on $K\times S^1\times S^1$
is equivalent to type IIA on $K$.

Let us discuss for instance the compactification
of F-theory on $K3$ \vf\ in the orbifold limits of $K3$
studied in \ref\seno{{A. Sen, ``F-theory and Orientifolds'',}
{\hep{9605150}}.}{\md},
from the viewpoint of U-duality alone.  We consider compactifying
from 10 dimensions down to 8 on a $T^2$.  Now the supercharges
belong to
$${\bf S}_8^\pm \otimes {\bf 1}_s^{\pm} \otimes {\bf 1}_s^{\pm}$$
where the first $\pm$ is correlated with the second one coming
from the $T^2$ going down from $10$ to $8$.
We will do a construction of type 2) in this setup.
We mod out by a discrete rotation
$(\omega , \omega ^{-1})$ where $\omega$ is a rotation
in the Kaluza-Klein $SO(2)$ and $\omega ^{-1}$ is rotation
in the U-group $SO(2)$
(as is well known the choices are very limited;
$\omega ^6$=1).  Then it is easy to see that there are
two invariant 8-dimensional spinors ${\bf S}_8^+, {\bf S}_8^-$.
By the chain of duality mentioned above upon compactification
on a circle this becomes equivalent to M-theory on an orbifold $K3$
given by $T^2\times T^2/(\omega , \omega ^{-1})$ and upon
compactification on an extra circle to type IIA on the same $K3$ orbifold.

\subsec{d=9}
In this case the $U$-duality group is still $SL(2)$ and this
is geometrized in the context of M-theory as noted above.

\subsec{d=8}
In this case the compact part of $U$-duality group is $SO(3)\times SO(2)$.
The $SO(2)$ part can be identified with the holonomy group of type IIB
compactification on $T^2$, and the $SO(3)$ can be viewed as a holonomy
group of compacitification of $M$-theory on $T^3$.  The
supercharges transform as
$$({\bf S}^+_8\otimes (2_s,1_s^+))+({\bf S}^-_8\otimes (2_s,1_s^-))$$
where $2_s$ denotes the spinor of $SO(3)$.  Note that we can
append a $T^2\times T^3$ to 8 dimensions and think about spinors
as being spinors on $R^8\times T^3\times T^2$, with the chirality
of the fermions being correlated between the 8-dimensional Minkowski
space and the internal $T^2$.
This is also in line with thinking
of the scalar moduli of the theory which are parameterized by
$$SL(2)\times SL(3)/ SO(2)\times SO(3)$$
as corresponding to flat metrics on $T^2\times T^3$, modulo an
overall volume factor on each torus (i.e. the moduli of `shapes' on each
torus).
We thus have a 13 dimensional
theory whose compactifications involves manifolds 
of $T^2\times T^3$ fibers.
Let us call this S-theory (we do not know the connection of this theory
to a 13-dimensional theory called S-theory in
\ref\bars{{I. Bars, ``S-Theory'',} {\hep{9607112}};
{``Algebraic Structure of S-Theory'',} {\hep{9608061}}.}\ but
we will use the same notation--hopefully they are related!).

In order to talk about compactifications of this theory
it is useful to connect it to F-theory, M-theory and type IIA
compactifications in lower dimensions.  To do this, note that if
we go down one dimension the $SL(2)\times SL(3)$ is U-conjugate
to the duality visible in F-theory, $SL(2)$ from the
elliptic fiber and $SL(3)$ from compactification on an extra $T^3$.
In other words we can identify the $T^2$ and $T^3$ of the F-theory with
the $T^2\times T^3$ we started with in 8 dimensions.  
Upon this reduction
a 4-brane of S-theory wrapped around circle gets identified with the
three brane of F-theory.
Thus we learn that
S-theory on $K\times S^1$ is equivalent to F-theory on
$K$. The volume of the $T^3$ in F-theory is mapped to the inverse radius
of the circle $S^1$.
Of course
we can continue the chain of dualities upon compactifications on another
circle leading to M-theory on $K$ where now the inverse radius
of the extra circle is related to the volume of the $T^2$ fiber
in M-theory.  Continuing compactification on an extra circle will lead
finally to type IIA on $K$.

The main difficulty with constructing interesting compactifications
of S-theory is the fact that manifolds $K$ which admit
both a $T^2$ and $T^3$ fiber {\it in a non-trivial way}
and which preserve some number of supersymmetries is {\it very limited}.
In fact the only class, which is not already covered by other theories
(in which only part of the $T^2\times T^3$ structure is used) involves
compactification on $K=CY_3\times K3_e$ where $CY_3$ is a threefold
Calabi-Yau which admits $T^3$ fibers, and $K3_e$ denotes a $K3$
which admits elliptic fibration.

First we ask if there are CY 3-folds admitting $T^3$ fibration.
In fact a certain class of them was constructed in
\ref\vwt{{C. Vafa and E. Witten, ``On Orbifolds with
Discrete Torsion'',} {J. Geom. Phys.} {\bf 15} {(1985)} {189}
{\hep{9409188}}.},
where a concrete example of mirror symmetry was reduced to T-duality
of the fiber $T^3$.  The construction involves 
considering $T^3\times T^3$
and modding out by an $SO(3)$ subgroup acting on each of the two $T^3$'s.
This will clearly give a Calabi-Yau, because $SO(3)\subset SU(3)$.
In fact a much larger class of Calabi-Yau's apparently admit
$T^3$ fibration and this has been conjectured to be the
basis for mirror symmetry
\ref\syz{{A. Strominger, S-T. Yau and E. Zaslow, ``Mirror
Symmetry is T-Duality'',}
{\hep{9606040}}.}\ref\mo{{D. Morrison, ``The Geometry Underlying Mirror
Symmetry'',}
{\alg{9608006}}.}\ref\gr{M.Gross and P.M.H. Wilson,
 ``Mirror Symmetry via 3-tori for a Class of
Calabi-Yau Threefold'',{\alg{9608004}}.}.

Upon compactification of S-theory on this 10 dimensional manifold, we
are down to 3-dimensions.  This theory has $N=2$ supersymmetry
in 3 dimensions (i.e. the same as reduction of $N=1$ supersymmetry
from 4 dimensions).  Upon compactification on an extra circle
this is dual to F-theory on $CY_3\times K3_e$.  Since
F-theory on $K3_e$ is dual to heterotic strings on $T^2$,
this means that in 2 dimensions we have a duality with heterotic
strings on $CY_3\times T^2$.  We can thus push this up one
step and obtain duality between heterotic strings on $CY_3\times S^1$
and S-theory on $CY_3\times K3_e$.

Note that for the heterotic compactifications on $CY_3$
we need to turn on 5-branes or put instantons.  Similarly
as was noted in
\ref\svw{{S. Sethi, C. Vafa and E. Witten, ``Constaints on Low
Dimensional String Compactifications'',} {\hep{9606122}}.}\
for F-theory
compactification we put appropriate configuration of 
3-branes for cancellation
of anomalies.  Similarly for S-theory we should put appropriate
configuration of 4-branes.

\subsec{d=7}
In this case the U-duality group is $SL(5)$ and the spacetime
supersymmetry charges belong to
the spinor of $SO(5)$.  We can append the 7-dimensional space
with a 5-dimensional torus $T^5$, whose shape is free to change
but its size is not dynamical, giving us the moduli space
$SL(5)/SO(5)$ which is the moduli of scalars in 7-dimensions.
Let us call this 12 dimensional thoery F'.
If we only use compactifications which factors in the form
of $T^2\times T^3$ we will simply be getting the F-theory
vacua.  If we use a $T^4$ fibration of it, we will be
getting M-theory vacua.  New vacua will arise if we use
the full $T^5$.  If we compactify on an extra circle the $SO(5)$
we have is conjugate to the $SO(5)$ of M-theory compactification
on $T^5$ where the volume of $T^5$ in M-theory
is identified with the inverse
radius of the circle we use for compactification.
  This implies that if we consider compactifications on
F'-theory on
$K$ which is a manifold admitting $T^5$ fibration,
upon further compactification on a circle we obtain
M-theory on $K$.

Again, unfortunately as in the case of S-theory
there are very limited possibilites of fully using
the whole $T^5$ structure to get new vacua
which preserve supersymmetry.  In fact the only new
class we are aware of involves compactifications
on CY 5-folds which admit $T^5$ fibrations.  These
will give rise to $(1, 1)$ supersymmetries  in $d=2$.  Upon
compactification to $d=1$ they are dual to M-theory
on the same CY 5-fold. To compare, we note that the 
compactification of F-theory on the same CY 5-fold leads to 
a $(2, 0)$ supersymmetric theory in $d=2$. 

\subsec{d$\leq $6}
As should be clear from the above discussion
the possibilities of constructing new vacua
using U-dualities are very limited.
This will also be the case (and in fact more true)
when we come down to lower dimensions.

For concreteness let us consider the case of $d=6$.
In this case the U-duality group is $SO(5,5)$ and the
compact subgroup is $SO(5)\times SO(5)$.  We can
append a $T^5\times T^5$ to the space; however now
the supersymmetry charge is not a spinor of $SO(5)\times SO(5)$;
rather it is a direct sum of spinors of each $SO(5)$:
$$[S_6\otimes (4,1)]\oplus [ S_6'\otimes (1,4)]$$
This implies that in studying the number of supersymmetries
we have, we need to study each $T^5$ fiber separately.  Note
however that the moduli of scalars is not $SL(5)\times SL(5)/SO(5)\times
SO(5)$, so the moduli of this theory is not the same as arbitrary metrics
up to scale on  $T^5\times T^5$.  It is more like the moduli of Narain
compacitifations on $T^5$.  In fact if we compactify further on a circle
it is U-conjugate to the $SO(5,5)$ T-duality of type II
on $T^5$.
If we consider the diagonal $SO(5)$, i.e. if
we consider the two $T^5$ fibers being the same it can be identified
with the $SO(5)$ holonomy of M-theory on $T^5$.  If we excite only the
$SO(4,4)$ fields, then this can be identified
with the T-duality in 6-dimensions and
can be related to asymmetric orbifold constructions
\ref\nsv{{K. Narain, M. Sarmadi and C. Vafa, ``Asymmetric
Orbifolds'',} {\np{288}{1987}{551}}.}.
As an aside let us note
an interesting asymmetric orbifold which acts on $T^4\times T^4$:
If we consider compactifications of type II on $T^4$, with
Narain lattice corresponding to the SO(8) group (where
left- and right-movers are $SO(8)\times SO(8)$ weight vectors
with difference in root lattice), and mod out by an overall
reflection on the left-movers and a translation by a fundamental
weight $(1,0,0,0)$ on the right-movers, we obtain a theory
in 6 dimension with $(4,2)$ supersymmetry (which in four
dimensional terms has $N=6$ supersymmetry).
Note that by construction all the states in
the twisted sector are massive.  All the moduli
consist of 4 scalars from R-R sector and 1 scalar, the dilaton,
from the NS-NS sector.
 This
theory has $SO(5,1)/SO(5)$ moduli space \jul ,
and we conjecture that it also has an $SO(5,1;Z)$ U-duality
symmetry which commutes with the element $g\in SO(5,5;{ \bf Z})$ used
in the orbifold construction.

Note that even though it is quite easy to construct
this vacuum in string perturbation theory
as an asymmetric orbifold it is not possible
to give a geometric construction of it starting
from either M-theory or F-theory.  This example
emphasizes how powerful string theory methods are,
despite the appearance of more abstract theories such
as M-theory or F-theory.

The connection of this U-duality group with the T-duality
in one lower dimension is also a good guide for constructing
vacua. In particular if we use the $SO(5,5)$ T-duality in construction
of some vacuum for string theory then there is a strong coupling
limit of it, where the theory grows one extra dimension and
is equivalent to making use of this 6-dimensional U-duality group.
Even so, the possibilities are very limited to get many new
vacua which have no perturbative string equivalent,
because in order to get something new we have to use
the full $SO(5)$ holonomy, and this is not easy to do
if one is also interested in preserving supersymmetries.
The possibilities are thus rather limited.

We would like to thank
K. Ray, V. Sadov, A. Sen and P. Townsend for useful discussions.

The research of C.V. was supported in part by NSF grant PHY-92-18167.

 \listrefs
\end